\documentclass[runningheads]{llncs}
\usepackage[utf8]{inputenc}
\usepackage[spanish]{babel}

\usepackage{graphicx}

\begin{document}

\title{Los sensores basados en dispositivos micromecánicos:
  laboratorios móviles al servicio de la enseñanza de las ciencias
  experimentales}

\titlerunning{Los sensores basados en dispositivos micromecánicos}

\author{Martin Monteiro\inst{1,2} \and
Cecilia Stari\inst{1} \and
Cecilia Cabeza\inst{1} \and Arturo C. Martí\inst{1}
}
\authorrunning{M. Monteiro et al.}

\institute{Instituto de Física, Universidad de la República, Uruguay \and
Universidad ORT Uruguay  \\
\url{http://smarterphysics.blogspot.com} \\
\email{marti@fisica.edu.uy}}
\maketitle              
\begin{abstract}
En este trabajo discutimos el uso de los sensores incorporados en los dispositivos móviles como posibles laboratorios ambulantes al servicio de la enseñanza de las ciencias experimentales. Los dispositivos móviles, smartphones, tabletas, laptops, tarjetas microbit, constituyen un recurso para realizar mediciones del mundo físico, ya que poseen un conjunto de sensores incorporados que permiten medir posición, velocidad lineal, velocidad angular, aceleración, presión, sonido, color, campo magnético, proximidad o luminosidad, entre otras. Asimismo, estos dispositivos han mejorado notablemente las prestaciones de sus cámaras de vídeo, permitiendo realizar fácilmente filmaciones a alta velocidad y alta resolución. Dada su portabilidad es posible trabajar en el propio laboratorio o en otros ámbitos como como un gimnasio, un parque o su propia casa, trascendiendo el ámbito tradicional del laboratorio. En general, las mediciones obtenidas pueden ser analizadas en el propio dispositivo o subidas a la nube para ser analizadas posteriormente.
\end{abstract}
\begin{abstract}
In this paper we discuss the use of sensors incorporated in mobile devices as possible mobile laboratories at the service of teaching experimental sciences. Mobile devices, smartphones, tablets, laptops, microbit cards, are a resource for making measurements of the physical world, since they have a set of built-in sensors that allow to measure position, linear speed, angular speed, acceleration, pressure, sound, color, magnetic field, proximity or luminosity, among others. Also, these devices have greatly improved the performance of their video cameras, allowing to easily shoot at high speed and high resolution. Given its portability it is possible to work in the laboratory itself or in other areas such as a gym, a park or your own home, transcending the traditional scope of the laboratory. In general, the measurements obtained can be analyzed in the device itself or uploaded to the cloud to be analyzed later.
\end{abstract}


\section{Introducción}

A pesar que los usuarios no siempre están al tanto, los dispositivos móviles, tablets y smartphones en general, disponen de un conjunto de pequeños sensores (cámara, micrófono, acelerómetro, GPS, entre otros) de última generación basados en dispositivos micromecánicos. Los fabricantes incluyen estos dispositivos con propósitos diversos, desactivar la pantalla táctil, regular el brillo, geolocalizar el dispositivo, rotar la pantalla como apaisada o retrato de acuerdo a la orientación del dispositivo. Es muy sabido en el público en general y en los actores educativos en particular, que estos sensores, potencialmente, transforman los dispositivos móviles en poderosas herramientas científicas con capacidad para, por ejemplo, generar y/o analizar sonido, luz, movimiento o ser usados como microscopios o espectroscopios. Gracias a la amplísima disponibilidad entres estudiantes y docentes, su portabilidad y facilidad de uso, estos dispositivos constituyen auténticos laboratorios ambulantes al servicio de la enseñanza de las ciencias naturales y en especial de la física.

\section{Los sensores basados en dispositivos micromecánicos}

Un sensor es un dispositivo capaz de medir una variable física o química y transformarla en una variable eléctrica. Los sensores presentes en los teléfonos inteligentes pueden ser de muy diversa naturaleza. Dado que la definición de sensor es muy general podemos pensar incluso el micrófono o la cámara digital como una clase particular de sensor. Una lista de los sensores más comunes y útiles para nuestros propósitos incluye a:

    • sensor de aceleración
    • sensor de rotación o giróscopo
    • micrófono
    • sensor de campo magnético
    • sensor de luz ambiente
    • sensor de proximidad,
      
pero existen otros, no tan frecuentes en los dispositivos móviles, como ser sensores de humedad, temperatura o presión. 

El sensor de aceleración o acelerómetro es uno de los de mayor utilidad en los experimentos de mecánica. En líneas generales podemos decir que un sensor de aceleración es una masa montada sobre un sistema de resortes tales que si la masa se acelera en una dirección uno de los resortes se comprime. A partir de la variación en la longitud del resorte, luego de su calibración, se puede determinar la aceleración. Estrictamente estos sensores no miden aceleración sino fuerzas. En la práctica, los teléfonos inteligentes usan sistemas más sofisticados basados en cerámicas piezoeléctricas o capacitores variables.

Diversos programas o aplicaciones permiten registrar los valores medidos por los sensores. En particular podemos mencionar \textit{Androsensor},o \textit{Physics Toolbox Suite}, otras posibilidades disponibles son \textit{Sensor Kinetics, Z-Device Tes}t. Habitualmente los sensores miden todas las componentes de las magnitudes vectoriales según los 3 ejes, \textit{x,y,z} orientados como si estuvieran dibujados sobre la pantalla del celular. Una vez registrados los datos es posible descargarlos en una computadora y analizarlos utilizando un programa apropiado.

\section{¿Cómo podemos usar los dispositivos móviles para hacer experimentos científicos?}

En la literatura especializada se han publicado diversas actividades y experiencias de Astronomía, Biología, Física y  Química basados en la utilización de los sensores que traen incorporados la mayoría de los dispositivos móviles ya sea tabletas o teléfonos inteligentes. Muchas veces, cuando dictamos charlas o talleres, los participantes se sorprenden cuando explicamos que tienen estas herramientas al alcance de la mano y más se asombran cuando les mostramos tan solo una parte del espectro de actividades que pueden realizar. 

Estas potencialidades que se derivan del uso de los sensores incorporados en los dispositivos móviles no están agotadas ni mucho menos. Dado que los fabricantes de los dispositivos no incorporan los sensores, por ejemplo el acelerómetro o el GPS, con fines educativos sino comerciales, la programación de las aplicaciones y el diseño de los experimentos es un tema delicado. La información disponible, por ejemplo, sobre sensibilidad, rango, alcance, frecuencia de muestreo, flujo de datos, es parcial y fragmentaria. En este sentido, la pregunta de cómo usar los sensores incorporados de los dispositivos móviles en experimentos de ciencias cobra especial relevancia. 

Nuestro grupo realiza numerosos aportes en la materia con diversas experiencias que apuntan a los aspectos antes reseñados: uso de los sensores, incluyendo el uso simultáneo de varios de ellos; reducción de los costos; realización de experimentos en lugares poco convencionales (parques, gimnasios, medios de transporte). Nuestros trabajos han sido publicados revistas arbitradas internacionales (ver http://smarterphysics.blogspot.com). 

Como mencionamos antes son numerosos los usos que se le puede dar a los sensores incorporados en los dispositivos móviles. Citamos a modo de ejemplos algunos de los posibles:

\begin{itemize}
 \item de la cámara para construir un microscopio casero \cite{hergemoller2017smartphone}
 \item de los sensores de ubicación y giro para determinar la orientación del celular e identificar un objeto celeste en el cielo nocturno \cite{schellenberg2015revisiting}.
 \item de la luz de la pantalla para estudiar los colores y su formación \cite{thoms2013color}.
 \item de la cámara junto a una pequeña red de difracción para estudiar espectros y fluorescencia \cite{polak2016easily}.
 \item de la cámara para estudiar realidad aumentada y virtual \cite{yuen2011augmented}.
 \item del micrófono y del parlante para estudiar sonidos en diversas circunstancias \cite{monteiro2015measuring,monteiro2018bottle}
 \item del sensor de aceleración para estudiar experimentos clásicos \cite{monteiro2015atwood,Dauphin_Bouquet_2018} o con objetos cotidianos \cite{tornaria2014understanding}.
 \item de los sensores de aceleración y rotación combinados para estudiar movimientos de personas y objetos \cite{monteiro2014rotational,monteiro2014angular,MONTEIRO2015,Monteiro2015analisis,Monteiro2016solucion,vieyrafive} en diversas configuraciones como por ejemplo un riel de aire comprimido \cite{Gonzalez2018air}
 \item del sensor de luminosidad para estudiar las propiedades de la luz \cite{monteiro2017polarization}.
 \item para medir la velocidad de ascensores, drones o incluso peatones subiendo escaleras \cite{monteiro2016using}.
 \item del sensor de campo magnético para estudiar el magnetismo en la materia \cite{monteiro2017magnetic}.
\item del sensor de presión y del GPS para estudiar las propiedades de la atmósfera \cite{monteiro2016exploring}.
\item de los sensores de aceleración y giro para hacer diversas experiencias en un parque de atracciones
\cite{pendrill2018mathematics,pendrill2018pendulum}
\item  de la entrada de micrófono para transformar el dispositivo en un pequeño osciloscopio \cite{forinash2018smartphone}.

\end{itemize}

Estas actividades reflejan solamente una pequeña parte del espectro de todas las posibles que involucran algunos sensores incorporados en los dispositivos móviles. La gran mayoría de las referencias son recientes, del orden de 4 o 5 años como máximo, reflejando el hecho que estos trabajos surgieron a partir de la masiva introducción de una tecnología que abarca no solo el ámbito educativo sino que ha permeado en la vida cotidiana de un porcentaje grande de la población mundial. 

\section{Algunas experiencias utilizando teléfonos inteligentes}

\begin{figure}[h]
\includegraphics[width=0.478\textwidth]{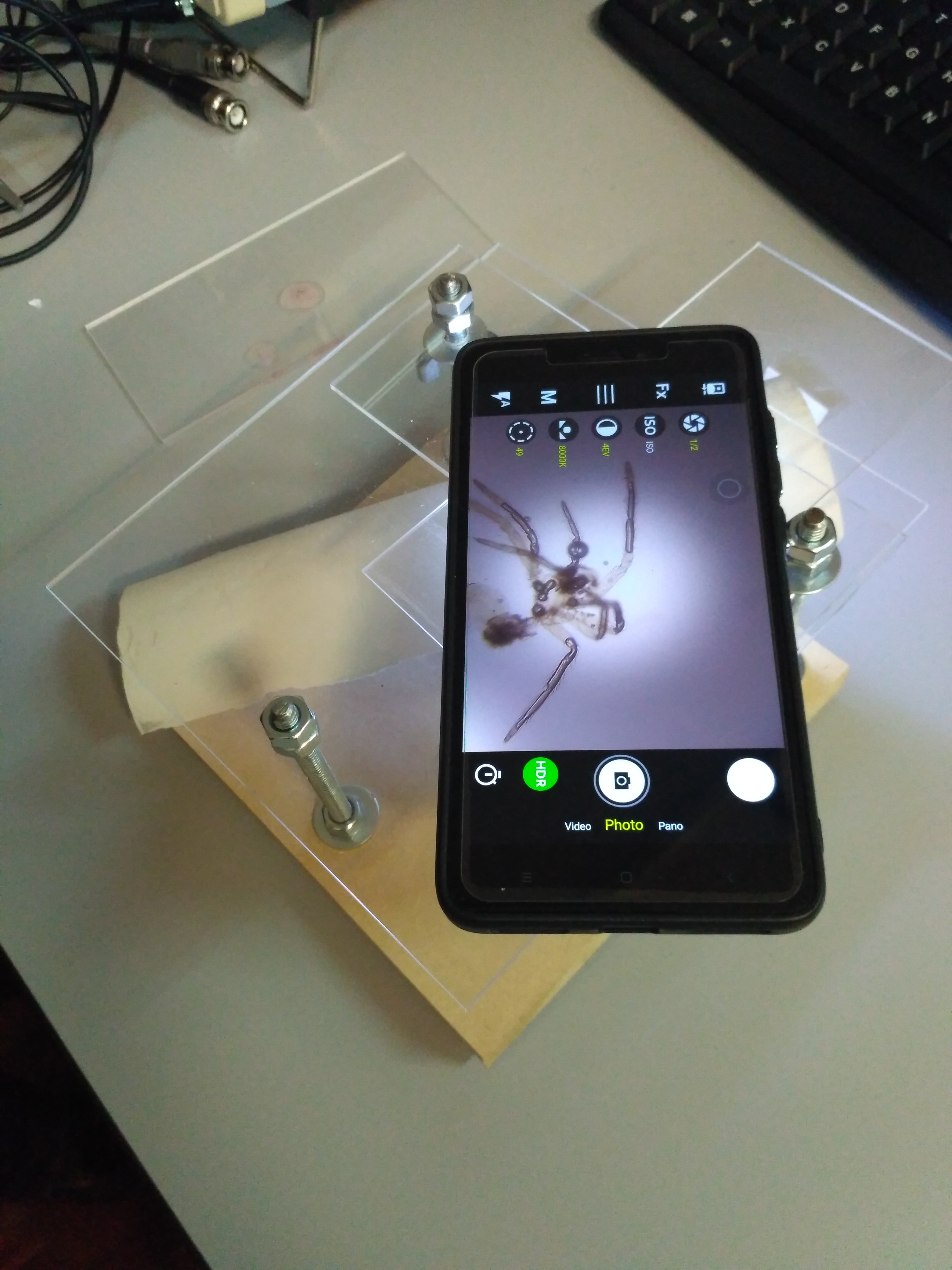}
\caption{Un smartphone transformado en microscopio sencillo usando materiales cotidianos. En la foto se aprecia como ejemplo una práctica realizada con una araña.
\label{expe4}}
\end{figure}

En esta sección revisamos algunos de los experimentos mencionados anteriormente. 
En primer lugar, en la figura \ref{expe4} mostramos un ejemplo de la utilización de un dispositivo móvil como pequeño microscopio.

\subsection{Un experimento de óptica}

En la sección anterior se listaron una diversidad de experimentos basados en sensores. 
Curiosamente, las experiencias relacionadas con aspectos tan cotidianos como la luz y en general
la óptica han sido menos explorados que otras ramas de la física, como la mecánica, las ondas o el electromagnetismo.

\begin{figure}[h]
\includegraphics[width=0.478\textwidth]{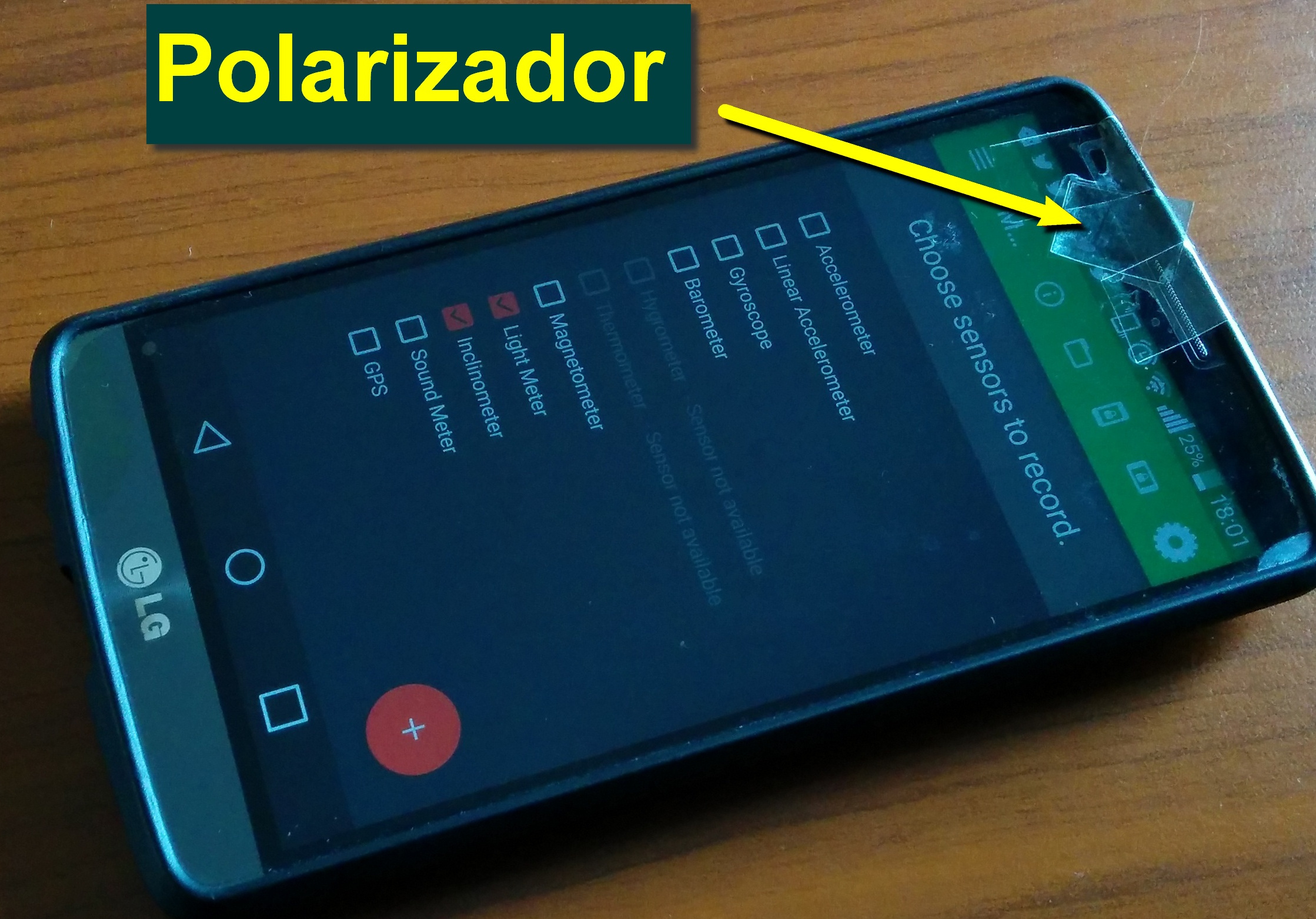}
\includegraphics[width=0.445\textwidth]{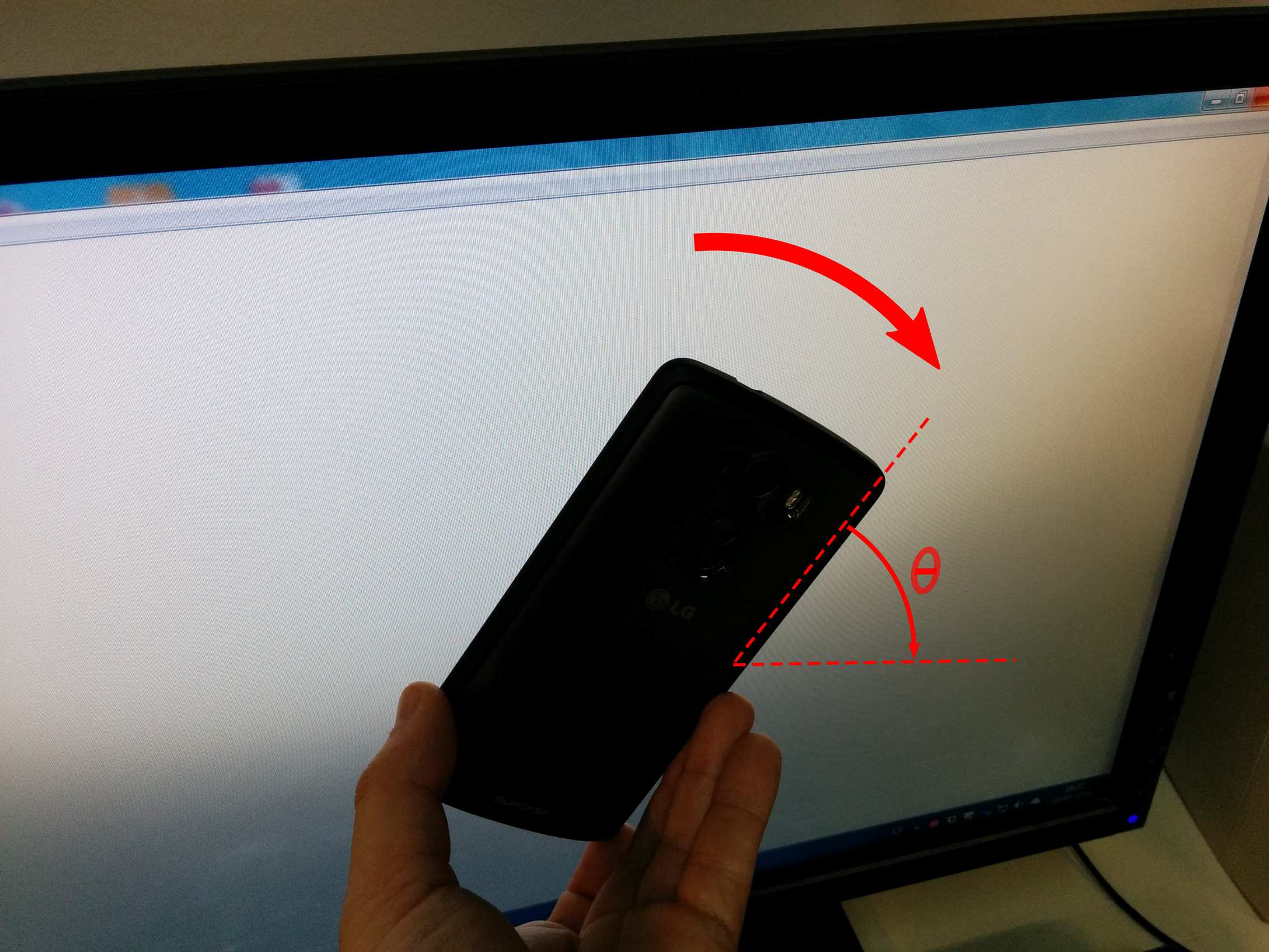}
\caption{En el panel derecho se observa como se coloca una lámina polarizadora (que puede obtenerse desarmando una calculadora vieja en desuso) sobre la cámara del dispositivo. Posteriormente, se hace uso de la luz brindada por un monitor y se gira el celular a la vez que se registran los datos de luz e inclinación con una aplicación adecuada (izquierda).
\label{expe1}}
\end{figure}

Mostramos a continuación una experiencia en la que utilizan las potencialidades de un dispositivo móvil 
 para verificar la experimentar con la polarización de la luz. Con este objetivo se emplea un monitor
 de computadora del tipo LCD cuya luz, se verifica fácilmente, está polarizada y  se mide 
 gracias al luxómetro (o medidor de luz ambiente) unido a un pequeño polarizador  mientras que el ángulo entre la polarización y el polarizador se mide por medio del sensor de orientación. En la figura \ref{expe1} se ilustra la configuración experimental.

 El uso simultáneo de estos dos sensores nos permite simplificar la configuración experimental y completar un conjunto de medidas en forma rápida. Los resultados experimentales de la intensidad de la luz en función del ángulo exhiben una excelente concordancia con la bien conocida ley de Malus.
Recalcamos  que los valores experimentales están en excelente acuerdo con los esperados, y concluimos que, gracias al uso simultáneo de dos sensores poco utilizados de un teléfono inteligente, es posible verificar la ley de Malus de una manera muy accesible para los estudiantes, promoviendo autonomía e involucramiento. Además, se pueden idear más experimentos utilizando teléfonos inteligentes y polarizadores, por ejemplo con placas de onda o retardadores (como placas de media onda y placas de cuarto de onda) o polarizadores circulares como los usados en algunas salas de cine en 3D o incluso con cinta de celofán de bajo costo para explorar birrefringencia.
 
\begin{figure}[h]
\includegraphics[width=0.85\textwidth]{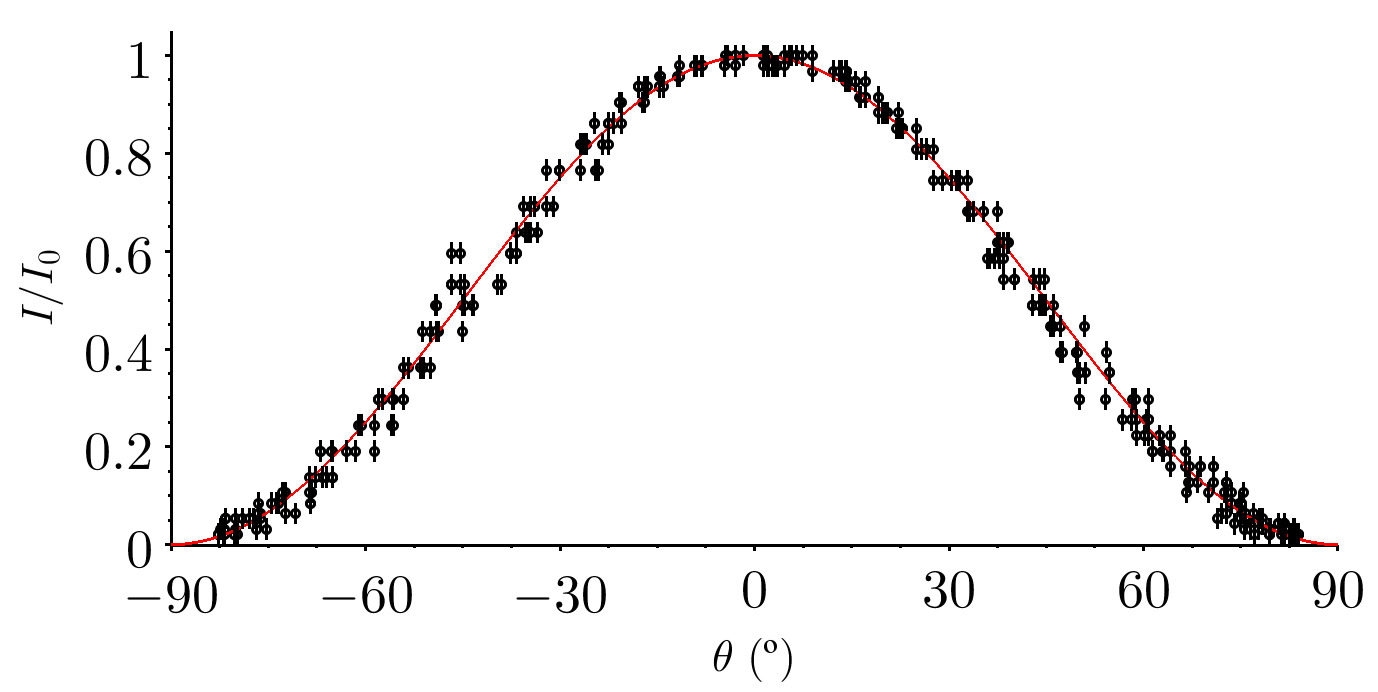}
\caption{Resultados experimentales (círculos), de intensidad de luz normalizada con respecto al máximo de luz transmitida en términos de 
la variable angular.  La Ley de Malus está representada por la línea continua. Como se observa la concordancia es  muy buena.
\label{expe2}}
\end{figure}

\subsection{Movimiento circular}

\begin{figure}[h]
\includegraphics[width=0.7\textwidth]{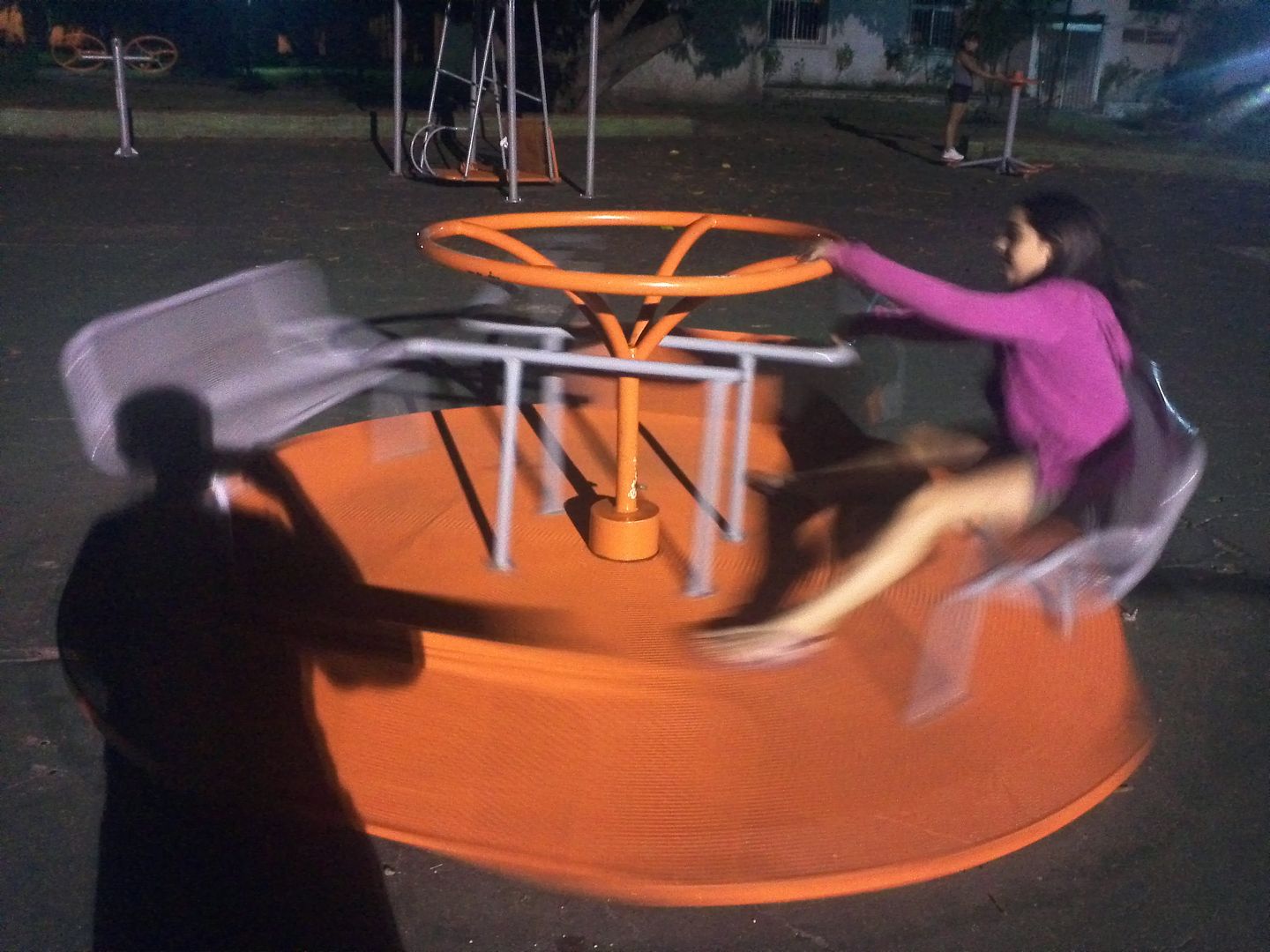}
\caption{Este experimento, donde se mide simultáneamente la velocidad angular y la aceleración centrípeta, puede ser realizado en una calesita de una plaza.
\label{expe3}}
\end{figure}

En esta experiencia se estudia el movimiento circular de un objeto. Para ello se utiliza el “teléfono inteligente” como cuerpo y se registrará el movimiento utilizando los sensores del acelerómetro y del giróscopo como se muestra en la figura \ref{expe3}.

Con los datos registrados se puede calcular la velocidad angular, el período del movimiento, la aceleración centrípeta y la variación del ángulo en función del tiempo. En esta práctica se verifica la relación lineal entre la velocidad tangencial y la velocidad angular
\begin{equation}
 a_c= \omega R.
\end{equation}
También es factible trabajar en las relaciones de conservación del momento angular y de la energía cinética.

\subsection{Péndulo simple}

En esta experiencia se estudia la dependencia del período de oscilación de un péndulo en relación a diferentes parámetros que pueden ser modificados, como la amplitud inicial de las oscilaciones o el largo del hilo. Se puede realizar la experiencia al menos por dos procedimientos diferentes. En un caso, se utilizará la cámara y se analizarán los resultados analizando el vídeo. En una segunda instancia, se medirá directamente con el uso de los sensores de rotación y aceleración de un celular. 

\subsection{El péndulo físico}

Un sistema físico paradigmático como es el péndulo físico se estudia usando los sensores de aceleración y rotación disponibles en los dispositivos  móviles. En nuestra experiencia un teléfono inteligente se fija en la parte exterior de una rueda de bicicleta y se pone en movimiento tanto en régimen de rotación (dando vueltas completas en una dirección) como de oscilación ya sea de pequeños o de grandes ángulos. Gracias a los sensores se obtienen medidas de aceleración y velocidad angular según diferentes ejes solidarios con el teléfono. El uso simultáneo de estos sensores permite además obtener trayectorias en el espacio de fases y visualizar las diferentes características del movimiento.

\section{¿Qué hace falta además de la tecnología?}

Diversas investigaciones muestran que la introducción de tecnología en sí misma, no alcanza para mejorar los logros de aprendizaje de los alumnos, si la misma no es utilizada junto a estrategias de enseñanza adecuadas. Desde hace algunos se ha mostrado que cuando se promueven estrategias activas los resultados son mejores en comparación con estrategias tradicionales basadas por ejemplo en clases magistrales (ver, por ejemplo, el trabajo pionero de Hake, \cite{hake1998interactive}). Las estrategias activas se basan en transformar el rol de la enseñanza, dándole un papel fundamental al estudiante, dejando de ser un receptor pasivo del conocimiento e involucrándose en las actividades en el aula. 

Una línea promisoria para fomentar una actitud activa en los estudiantes es el uso de los dispositivos móviles concebidos como una herramienta científica. Los mismos son otra puerta para realizar mediciones del mundo físico, ya que poseen un conjunto de sensores incorporados que permiten medir variables físicas del entorno. Asimismo, estos dispositivos han mejorado notablemente las prestaciones de sus cámaras de vídeo, permitiendo realizar fácilmente filmaciones excelente calidad. Dada su portabilidad es posible trabajar en el propio laboratorio o en otros ámbitos como como un gimnasio, un parque o su propia casa, trascendiendo el ámbito tradicional del laboratorio. En general, las mediciones obtenidas pueden ser analizadas en el propio dispositivo o subidas a la nube para ser analizadas posteriormente.

En general, los efectos del uso de los dispositivos móviles y sus sensores en el sistema educativos también son cuestiones abiertas y novedosas. Dado que son tecnologías muy recientes los estudios donde se analiza su impacto son todavía preliminares y escasos. Las investigaciones que apuntan a un mejor uso de los recursos también son escasas y las preguntas abiertas son muchas. ¿Se sienten los estudiantes más motivados a usar su propio dispositivo de uso personal qué otro aparato ajeno a su universo inmediato? La posibilidad dada por estos dispositivos de una retroalimentación muy inmediata, ¿afecta la actitud frente al aprendizaje? ¿Cómo se ven afectadas las cuestiones de privacidad? No solo el uso de los dispositivos móviles en sí plantea desafíos sino también la disponibilidad de grandes cantidades de datos y la posibilidad de analizarlos con poderosas herramientas es un aspecto removedor que vale la pena estudiar con detalle. Todas estas preguntas están abiertas y seguramente en los próximos años sean objeto de intenso debate.

\section{Conclusiones y perspectivas}

Como se desprende las consideraciones anteriores los dispositivos móviles ofrecen una variedad muy grande de posibilidades para experimentar en Ciencias Naturales. Queremos mencionar aquí una característica especialmente destacable que es la capacidad  para medir simultáneamente con varios sensores. Esto hecho constituye una gran ventaja ya que permite realizar una gran variedad de experimentos, incluso al aire libre, evitando la dependencia de instrumentos frágiles o no disponibles. En trabajos anteriores se propuso el uso simultáneo de dos sensores como el giroscopio y el acelerómetro para relacionar la velocidad angular, la energía de rotación, la aceleración centrípeta y tangencial \cite{MONTEIRO2015,Monteiro2015analisis}. En otro experimento, el sensor de presión y el GPS se utilizaron en sincronía para encontrar la relación entre la presión atmosférica y la altitud \cite{monteiro2016exploring}.
También el sensor de presión ha sido usado en conjunto con el acelerómetro \cite{monteiro2016using}.

La cantidad de dispositivos móviles vendidos en todo el mundo continua aumentando. Asimismo, los dispositivos vienen cada vez mejor equipados y con más sensores. La sociedad en general, pero especialmente los jóvenes, estamos acostumbrados a llevar el smartphone a todas partes. Estamos convencidos que los usos educativos de los dispositivos móviles también seguirán aumentando en los próximos años. El único límite para aprovechar las potencialidades es, como dijo alguien, nuestra propia imaginación.

%
%
\bibliographystyle{splncs04}
\bibliography{/home/arturo/Dropbox/bibtex/mybib}

\end{document}